\documentclass[twocolumn,aps]{revtex4}

\usepackage{amsmath}
\usepackage{amssymb}
\usepackage{graphicx}
\usepackage{bm}

\newcommand{\beq}{\begin{equation}}
\newcommand{\eeq}{\end{equation}}
\newcommand{\ben}{\begin{eqnarray}}
\newcommand{\een}{\end{eqnarray}}
\newcommand{\benn}{\begin{eqnarray*}}
\newcommand{\eenn}{\end{eqnarray*}}
\newcommand{\lap}{\nabla^2}
\newcommand{\lapi}{\nabla^{-2}}
\newcommand{\hz}{\hat{z}}
\newcommand{\bx}{\mathbf{x}}
\newcommand{\fcal}{\mathcal{F}}
\newcommand{\lcal}{\mathcal{L}}
\newcommand{\gcal}{\mathcal{G}}
\newcommand{\ncal}{\mathcal{N}}
\newcommand{\bp}{\mathbf{B}_p}
\newcommand{\jp}{j_{\parallel}}
\newcommand{\bxi}{\bm{\xi}}
\newcommand{\bqp}{\mathbf{Q}_{\perp}}

\newcommand{\bB}{\mathbf{B}}
\newcommand{\rot}{\nabla\times}
\newcommand{\bQ}{\mathbf{Q}}
\newcommand{\va}{v_A^2}
\newcommand{\hy}{\hat{y}}
\newcommand{\tbp}{\tilde{\mathbf{B}}_p}

\begin{document}

\title{Stability of compressible reduced magnetohydrodynamic equilibria\\ -- analogy with  magnetorotational instability}
\author{P. J. Morrison}
\email{morrison@physics.utexas.edu}
\affiliation{Department of Physics and Institute for Fusion Studies, University
  of Texas at Austin, Austin, TX 78712, USA}
\author{E. Tassi}
\email{tassi@cpt.univ-mrs.fr}
\affiliation{Centre de Physique Th\'eorique\\
Aix-Marseille Universite, CNRS, CPT\\
 UMR 7332, 13288 Marseille, France}
 \author{N. Tronko}
\email{nathalie.tronko@gmail.com}
\affiliation{York Plasma Institute, Physics Department\\
 University of York, YO10 5DD, York, UK}
\affiliation{Centre for Fusion, Space and Astrophysics,
 Physics Department\\
University of Warwick,  CV4 7AL, Coventry, UK}

\begin{abstract}

Stability analyses for equilibria of the compressible reduced magnetohydrodynamics (CRMHD) model are  carried out by means of the Energy-Casimir (EC) method. Stability results are compared with those obtained for ideal magnetohydrodynamics (MHD) from the classical $\delta W$ criterion.  An identification of the terms in the second variation of the free energy functional for CRMHD with those of  $\delta W$  is made:   two destabilizing effects present for CRMHD turn out to correspond to the kink and interchange instabilities in usual MHD, while the stabilizing roles of field line bending and compressibility are also identified in the reduced model.  Also, using the  EC method,  stability conditions in the presence of toroidal flow are obtained.  A formal analogy between CRMHD and a  reduced incompressible model for magnetized rotating disks,  due to Julien and Knobloch [EAS Pub.\  Series, {\bf 21}, 81 (2006)], is discovered. In  light of this analogy,  energy stability analysis  shows that the condition for magnetorotational instability (MRI) for the latter model,  corresponds to the condition for interchange instability in CRMHD, with the Coriolis term and shear velocity playing the roles of the curvature term and pressure gradient, respectively.  Using the EC method, stability conditions for the rotating disk model, for a large class of equilibria with possible non-uniform magnetic fields, are obtained. In particular, this shows it is possible for the MRI system to undergo, in addition to the MRI, another instability that is  analogous to the kink instability.  For vanishing magnetic field,   the Rayleigh hydrodynamical stability condition is recovered.

\end{abstract}
\maketitle

%%%%%%%%%%%%%%%%%%%%%%%%%%%%
%%%%%%%%%%%%%%%%%%%%%%%%%%%%
\section{Introduction}  \label{sec:intro}

The investigation of instabilities is of paramount importance for fusion and astrophysical plasmas.
For fusion plasmas, detecting the main instabilities is necessary to learn how plasma characteristics such as current, density,  and temperature should be optimized in order to improve plasma confinement.
For astrophysical plasmas, the identification of instability mechanisms is essential for understanding phenomena such as solar flares, coronal mass ejections, transport of angular momentum in accretion disks,  and several other major observable phenomena.  In the framework of the fluid description of plasmas, the use of the energy principle is one of the possible ways to investigate linear stability in the limit where dissipative processes are negligible.

In their early work, Bernstein et al.\ \cite{Ber58} provided the energy principle for the investigation of stability in ideal MHD. According to such principle, an ideal MHD equilibrium is stable if and only if the quadratic form $\delta W$ (more aptly called $\delta^2W$, but we follow tradition), representing the second variation of the potential energy, induced by variations of the field variables, is positive or zero, for all the allowable Lagrangian displacements of the fluid.
Further extensions of the energy principle, and application of it to particular geometries have been developed in the subsequent years (see, e.g.,  Ref.~\cite{Fre87}), and  methods based on the energy principle  are now  standard  for investigating plasma stability.

Alternatively, the EC method, an Eulerian stability method with an early antecedent in plasma physics \cite{Kru58}, evolved into a formalized and systematic procedure because of  the discovery of the noncanonical (Eulerian) Hamiltonian description of ideal MHD  in Ref.~\cite{Mor80}.  Of particular relevance to the present paper are a large number of works that followed  \cite{Mor80} detailing the  noncanonical Hamiltonian formulation of reduced models,  viz.,  reduced MHD \cite{Mor84,Mar84}, the Charney-Hasegawa-Mima equation \cite{Wei83}, the four-field model for tokamak dynamics \cite{Haz87}, models for collisionless reconnection \cite{Sch94,Tas08}, hybrid fluid-kinetic models \cite{Tro10},  and models for many other systems.  (See \cite{Mor82a, Mor98, Mor05} for review.)

Hamiltonian structure provides a natural means of obtaining sufficient conditions for stability, and this is generalized for the  Hamiltonian structure of fluid models for plasmas that are defined on  an infinite-dimensional phase-space and are  noncanonical.  The latter property gives rise to the existence of a special class of invariants for the model: the Casimirs. These are functionals of the observables, that  commute, through the Poisson bracket of the system, with any other functional of the observables.   The identification of the Casimirs for a system, provide the basis for the EC method (see, e.g.,  \cite{Haz84,Hol85,Mor86,Mor98} for a review).

According to the  EC method, stability is attained if the second variation $\delta^2 F$, where $F$ is a free energy  functional obtained as a linear combination of the Hamiltonian and the Casimirs, has a definite sign, when evaluated at the equilibrium of interest.   This method can be applied to the huge class of plasma models possessing noncanonical Hamiltonian structure.   It provides sufficient conditions for energy stability, a strong form of stability that  implies linear stability  (while the converse is in general not true) and  almost nonlinear stability.

Given the above considerations, the question naturally arises of whether there is a  connection between the stability results obtained from the EC method and those of the  $\delta W$  energy principle. One goal of this paper is to explicitly trace  a connection between the two methods in the context  of two reduced fluid models: the CRMHD model and the accretion disk model introduced  in Ref.~\cite{Jul06}, in their dissipationless limits.

CRMHD \cite{Mor84b}, was conceived to investigate tokamak dynamics in a simplified geometry, using a large aspect-ratio ordering. It generalizes reduced MHD by taking into account finite compressibility, which involves a coupling between parallel flow and pressure evolution. It accounts also for  effects due to magnetic field curvature. Its Hamiltonian structure was derived in Ref. \cite{Haz87}.

The accretion disk model, on the other hand, describes locally an incompressible rotating plasma in the presence of an imposed azimuthal sheared flow and a poloidal field. It has been derived, via an asymptotic expansion, in order to investigate, in a simplified setting, the MRI \cite{Vel59, Cha60, Ach73fl, Ach73rev}, which is believed to play a key role in the transport of angular momentum in accretion disks \cite{Bal91}. The Hamiltonian structure for this model, obtained here,  was not previously known.

In this paper, we carry out an analysis in the case of CRMHD, first. For a wide class of  equilibria, we are able to trace the connection between the two methods. In particular, the terms that can lead to indefiniteness in the sign of $\delta^2 F$,  actually correspond to the terms associated with the kink and interchange instabilities in $\delta W$. We identify also a correspondence between the stabilizing terms that appear in the expressions obtained from the two methods.

The analysis in the case of the accretion disk model follows essentially as a by-product of an identification  rule  that permits us to map one model into the other. This provides us with a convenient framework for achieving another  objective of this article, which is the identification of analogies between instabilities relevant for tokamaks and those for accretion disks. By taking advantage of the mapping, we are able to show that  the MRI model possesses a Hamiltonian structure. We apply then the EC method to this model and derive energy stability conditions. In addition to recovering the expected condition for MRI, we are able to see that such instability in the accretion disk model, plays the role of the interchange instability in CRMHD. Thanks to the unifying framework provided by the EC method and to the connection with the energy principle, we can establish further analogies, from which  emerge  quite strong similarities  between the two physical systems, providing an example of universality between instabilities in fusion and astrophysical plasmas.

The article is organized as follows.   In Sec.~\ref{sec:crmhd},  we recall the CRMHD model, briefly review  the noncanonical Hamiltonian description for continuous media, and then derive stability conditions for CRMHD using  the EC method. Next, stabilizing and destabilizing terms  of the EC method are compared to those of the $\delta W$  energy principle by mapping Lagrangian variations to Eulerian.  In Sec.~\ref{sec:analogy} we present the formal correspondence  between CRMHD and the MRI model. Section \ref{sec:mristab} is devoted to the stability analysis of the MRI model and to the identification of the similarities with CRMHD.  Finally, we conclude  in Sec.~\ref{sec:concl}.

%%%%%%%%%%%%%%%%%%%%%%%%%%%%
%%%%%%%%%%%%%%%%%%%%%%%%%%%%
\section{CRMHD}
 \label{sec:crmhd}

\begin{figure}[h!]
\centering
\includegraphics[width=8.5cm]{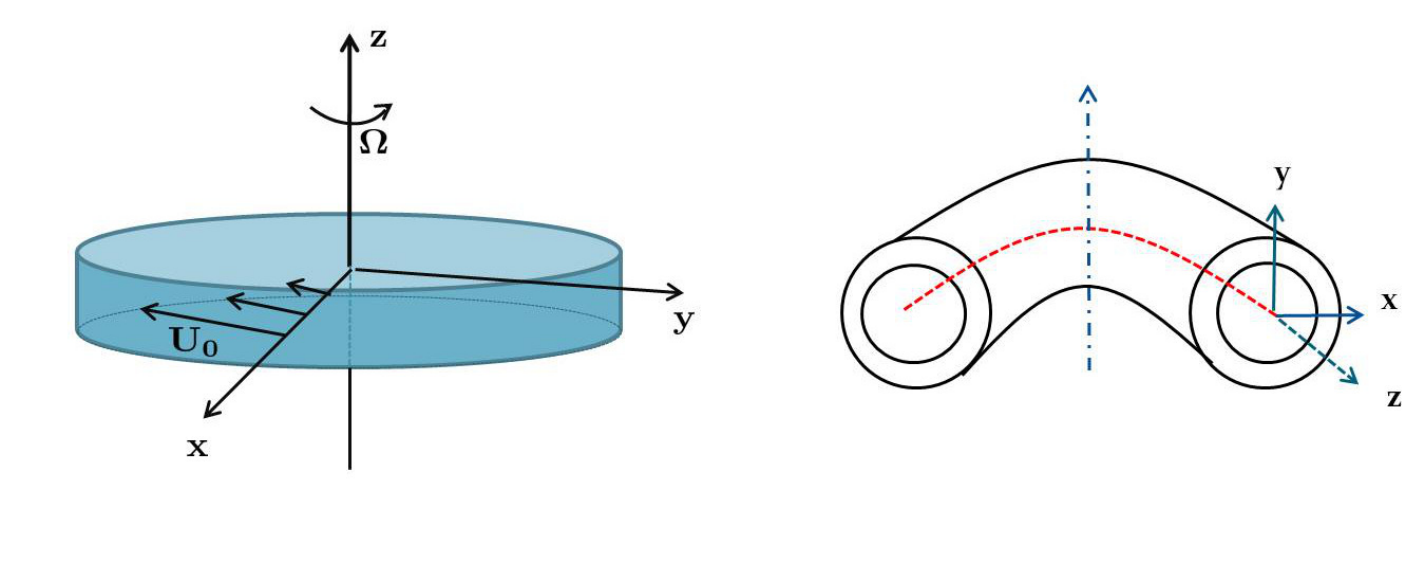}
\caption{Sketches of the geometry and  local coordinate system for the accretion disk model (left) and for CRMHD for  a tokamak (right).}
\label{fig1}
\end{figure}

In a Cartesian coordinate system $(x,y,z)$ as depicted in Fig. \ref{fig1}, the equations for CRMHD are given by
\ben
\frac{\partial \omega}{\partial t}&+& [\phi, \omega]_y+[\lap \psi , \psi]_y-2[p,h]_y=0, \label{c1}\\
\frac{\partial \psi}{\partial t}&+& [\phi,\psi]_y=0,  \label{c2} \\
\frac{\partial v}{\partial t}&+& [\phi , v]_y+[p, \psi]_y=0, \label{c3}\\
\frac{\partial p}{\partial t}&+& [\phi , p]_y+\beta [v, \psi]_y-2 \beta [h , \phi]_y=0
 \label{c4}.
\een
Equations (\ref{c1})--(\ref{c4}), describe the dynamics of a plasma in a magnetic field $\mathbf{B}=\varepsilon\nabla\psi \times \hz + [(1+\varepsilon x)^{-1} + \varepsilon \bar{b}]\hz + \mathcal{O}(\varepsilon^2)$, with $\varepsilon=a/R_0\ll 1$ being the ratio of the characteristic length scale of the poloidal plane, $a$, with the major radius of the magnetic axis, $R_0$,  and $\bar{b}$ accounting for diamagnetic corrections. The fields $v$ and $p$ represent the parallel ion velocity and the electron pressure, respectively, whereas $\omega=\nabla^2\phi$ with $\phi$ being the electrostatic potential.  All the fields of the model are independent of the coordinate $z$.  The quantity $h\equiv x$ accounts for the effects of magnetic field curvature, while  the parameter $\beta$ is the ratio of the background electron fluid pressure and the magnetic pressure. Finally, the quantity $[f,g]_y=\hz \cdot \nabla f \times \nabla g$ is  the canonical bracket, with respect to the variables $x$ and $y$, between two functions $f$ and $g$.

Equation  (\ref{c1}) is the equation for the evolution of the $z$ component of the vorticity, $\omega$,  associated with the $\mathbf{E}\times\mathbf{B}$ flow, while  Eq.~(\ref{c2}) expresses the frozen-in condition for the poloidal magnetic flux $\psi$.    Equation (\ref{c3}) governs the parallel ion dynamics and Eq.~(\ref{c4}) determines the electron pressure evolution.

%%%%%%%%%%%%%%%%%%%%%%%%%%%%
\subsection{CRMHD Hamiltonian structure}

The CRMHD system of Eqs.~(\ref{c1})--(\ref{c4}) is known to possess a noncanonical Hamiltonian structure \cite{Haz87,Thi98}, which we review here.  To this end we first recall  a few basic notions about infinite-dimensional Hamiltonian systems (for further details see, e.g.,  Ref. \cite{Mor98}).

Given a field theory $\dot{\chi}^i(\bx)=F^i(\chi^1, \cdots , \chi^N)$, for $N$ fields $(\chi^1(\bx), \cdots , \chi^N ( \bx))$, evolving according to the vector fields $F^i$, we say that such system is a Hamiltonian system if a Hamiltonian functional $H(\chi^1, \cdots \chi^N)$ and a Poisson bracket $\{ , \}$ exist, such that the field equations can be written in the form
\ben  \label{hamsys}
\dot{\chi}^i=\{\chi^i , H \}, \qquad i=1, \cdots ,N,
\een
where the  Poisson bracket must be bilinear, antisymmetric,  and  satisfy the Leibniz and Jacobi identities.

For continuous media,  such as plasmas described in terms of Eulerian variables, the case for CRMHD, Poisson brackets typically possess a noncanonical structure. In these cases there exist Casimir functionals, that is,  functionals $C(\chi^1, \cdots , \chi^N)$  that satisfy $\{C,F \}=0$, for all functionals $F(\chi^1, \cdots , \chi^N)$.  Because,  $\dot{C}=\{C,H\}=0$, for  any Hamiltonian, Casimirs are invariants of the dynamics.

For the CRMHD, the Hamiltonian functional is given by
\ben  \label{hcrmhd}
H=\frac{1}{2}\int dxdy \left( \vert \nabla \phi \vert^2 + v^2 + \vert \nabla \psi \vert^2 + {p^2}/{\beta} \right).
\een
All the terms in the Hamiltonian are amenable to a physical interpretation: the first two terms represent the kinetic energy associated with the $\mathbf{E}\times\mathbf{B}$ flow and with the toroidal flow, respectively. The third term is the poloidal magnetic energy, whereas the last term accounts for the internal energy of the electron fluid.

The noncanonical Poisson bracket for CRMHD, on the other hand, is given by
\ben
\{F,G\}&=&\int\!dxdy \Big(\omega [F_{\omega}, G_{\omega}]_y\nonumber\\
&+&v([F_{\omega}, G_v]_y+[F_v , G_{\omega}]_y) \nonumber\\
 &+&   (p+2 \beta h)([F_{\omega}, G_p]_y+[F_p , G_{\omega}]_y) \nonumber\\
 &+&    \psi([F_{\omega}, G_{\psi}]_y +[F_{\psi}, G_{\omega}]_y\nonumber\\
&{\ }&  - \beta [F_p ,G_v]_y-\beta [F_v, G_p]_y)
 \Big)\,,
 \label{bcrmhd}
\een
where subscripts of the functionals indicate functional derivatives.
The operation (\ref{bcrmhd}) can be shown to satisfy all the properties of a Poisson bracket, provided that surface terms vanish when integrating by parts. This can be accomplished, for instance, on a bounded domain, imposing that the fields and the functional derivatives are periodic in the $y$ direction and satisfy vanishing Dirichlet boundary conditions along $x$.
The Hamiltonian (\ref{hcrmhd}) and the bracket (\ref{bcrmhd}), combined according to (\ref{hamsys}), yield namely the CRMHD system (\ref{c1})-(\ref{c4}).
Casimir functionals of the bracket (\ref{bcrmhd}), on the other hand, are given by
\ben
C_1&=&\int dxdy\, \mathcal{F}(\psi),  \nonumber\\
C_2&=&\int dxdy \ v\ \mathcal{N}(\psi), \nonumber\\
C_3&=&\int dxdy\, \mathcal{L}(\psi)\left({p}/{\beta}+2h\right), \nonumber\\
C_4&=&\int dxdy\, \left(\omega\ \mathcal{G}(\psi)-v\  \mathcal{G}'(\psi)\, \left({p}/{\beta}+2h\right)\right),
 \label{ccrmhd}
\een
where $\fcal$, $\ncal$, $\lcal$ and $\gcal$ are arbitrary functions, and the prime denotes derivative with respect to the argument of the function.  This set of  Casimirs  is close to but not equivalent to that for unreduced MHD with various symmetries \cite{amp0,amp1}.  Because of the infinite number of invariants of Eqs.~(\ref{ccrmhd}) the dynamics of CRMHD is constrained.  In particular, the existence of the family of Casimirs $C_1$ reflects the constraint of the frozen-in condition for the poloidal magnetic flux. The families $C_2$ and $C_4$, on the other hand, are remnants  of the cross-helicity $\int d^3 x\ \mathbf{v}\cdot\mathbf{B}$, which is a Casimir for isentropic ideal MHD (see Ref.~\cite{padhye}), whereas $C_3$ derives from the conservation of magnetic helicity of  ideal MHD \cite{Thi98}. Such relationships  with the Casimirs of full ideal MHD make use of the proportionality between $p$ and $\bar{b}$, which is imposed when deriving CRMHD from the fluid plasma description  (see, e.g.,  Ref.~\cite{Haz85}).

%%%%%%%%%%%%%%%%%%%%%%%%%%%%
\subsection{CRMHD equilibria and energy stability}

One of the advantages provided by the Hamiltonian structure of a model is the possibility of using specific, well-developed  methods for Hamiltonian systems;  in particular, methods that have been developed for determining stability conditions.
For noncanonical systems, such as CRMHD, the EC method is particularly convenient, because it can take advantage of  the infinite number of Casimirs possessed by the model. The EC method applies to equilibria that are extremals of a {\it free energy functional} $F$,  which is a linear combination of the Hamiltonian and the Casimirs. For such equilibria, the method can provide sufficient conditions for energy stability -- stability  that arises from  the existence of a constant of motion for the system, whose second variation has a definite sign, when evaluated at the equilibrium of interest. Because, the Hamiltonian and the Casimirs are constants of motion, then clearly $F$  is also a constant of motion. Therefore, for noncanonical Hamiltonian systems the free energy functional provides a natural constant of motion from which one can determine energy stability.   We remark that this type of stability implies linear stability, but the converse is not true in general.  Also, we remark that this kind of stability almost assures nonlinear stability, but in general nonlinear stability does not imply linear stability.

The EC method evolved out of  early plasma work \cite{Kru58} and there is now considerable prior application to plasma physics problems, as can be found in Refs.~\cite{Haz84,Hol85,Mor86,Dag05,Kha05,Tas11}.

In order to apply  the EC method to CRMHD, we define first the free energy functional, which for CRMHD, in its most general form, reads
\beq \label{fenc}
F=H+C_1 +C_2 +C_3 +C_4.
\eeq
Equilibria to which the EC method can be applied are extremals of (\ref{fenc}), i.e.,  those at which the first variation
\ben
\delta F&=&\int dxdy \bigg[
\Big(-\lapi \omega + \gcal (\psi) \Big)
\delta \omega
\nonumber\\
&+&  \Big( - \lap \psi + \fcal ' (\psi) + v\ \ncal ' (\psi)+\left({p}/{\beta}+2h \right) \lcal'(\psi)
\nonumber\\
&{\ }&   +\  \omega\,  \gcal ' (\psi) - v\  \gcal '' (\psi) \left({p}/{\beta} + 2h\right)\Big)
\delta \psi
\nonumber\\
&+&  \Big( v\  + \ncal (\psi) -\gcal ' (\psi) \left( {p}/{\beta} + 2h \right)\Big)
\delta v
\nonumber\\
&+& \Big( p + \lcal (\psi) - v\ \gcal ' (\psi) \Big) \delta p /\beta
\bigg]
\een
vanishes. Because of the arbitrariness of the variations $\delta \omega$, $\delta \psi$, $\delta p$ and $\delta v$, such equilibria must be the solutions for $\phi$, $\psi$, $p$ and $v$, of the system
\ben
\phi&=&\gcal (\psi),
\label{ec1} \\
\lap \psi&=&\fcal ' (\psi) + v\ \ncal ' (\psi) + \left({p}/{\beta} + 2h\right)\lcal ' (\psi)
\nonumber\\
&{\ }&  + \omega\ \gcal ' (\psi) - v\ \gcal '' (\psi) \left({p}/{\beta} + 2h\right),
\label{ec2} \\
v&=& -\ncal (\psi) + \gcal ' (\psi) \left({p}/{\beta} + 2h \right),
 \label{ec3} \\
p&=&-\lcal (\psi) +v\ \gcal ' (\psi),
 \label{ec4}
\een
 for  choices of $\gcal$, $\fcal$, $\ncal$ and $\lcal$.

 We remark first that the equilibrium equation (\ref{ec1}) expresses the condition that, at equilibrium, the poloidal flow is a poloidal magnetic flux function. In the following, however, we will consider only the case $\gcal\equiv0$, corresponding to the absence of poloidal flow, which is a  standard case treated  in  linear stability analyses of ideal MHD. However, with  the following transformation to a {\it dynaflux} function $\chi$:
 \beq
 \chi=\int^{\psi} \!\!d\bar\psi \, \sqrt{1- \gcal'^2(\bar\psi)}\,,
 \eeq
which is valid for  sub-Alfv\'enic flows ($\gcal'^2<1$), equilibria with poloidal flow can be mapped into ones without.  This transformation was first noted in Ref.~\cite{morrison86}  and generalized in Ref.~\cite{tasso}.  We will not pursue this further here, but emphasize that for every equilibrium without poloidal flow  there exist a family of equilibria determined by $\gcal$  with poloidal flow.

If $\gcal\equiv0$, then Eqs.~(\ref{ec3}) and (\ref{ec4}) imply that  toroidal flow and  pressure are constant on poloidal magnetic surfaces. Inserting (\ref{ec3}) and (\ref{ec4}) into (\ref{ec2}), yields
\beq  \label{gs}
 \lap \psi =\fcal ' (\psi) - v'(\psi) v(\psi) - p'(\psi) \big({p (\psi) }/{\beta}+ 2h\big),
 \eeq
 which can be seen as a generalized Grad-Shafranov equation for the magnetic flux function $\psi$.
 Once this partial differential equation is solved for  $\psi$, the corresponding solutions for the fields $p$ and $v$ can be readily obtained from the specific choices  of the free functions $\lcal$ and $\ncal$.

 When investigating stability for a reduced  model, such as CRMHD, it is natural to ask how the stability properties of the reduced model square with  those of the complete MHD equations.  This can help in interpreting the stability conditions for the reduced model and in seeing what features have possibly been lost because of the reduction. In the following we compare the stability analysis for CRMHD carried out by means of the EC method, with the classical energy principle analysis \cite{Ber58}, based on the $\delta W$ criterion, for the complete ideal MHD system with no equilibrium flow.

  In order to facilitate this comparison, we   consider first the special case where $\ncal\equiv0$. Together with $\gcal\equiv0$, this assumption implies static equilibrium for CRMHD, i.e., equilibria with neither poloidal nor toroidal flow. For this case, the second variation of the free energy functional reads
 \ben  \label{2fc1}
 \delta^2 F&=&\int dxdy \bigg[
  \vert \nabla \delta \phi \vert^2 + \vert \nabla \delta \psi \vert^2
  +\vert \delta v\vert^2 + {\vert \delta p \vert^2}/{\beta} \nonumber\\
 &+&\left( \fcal '' (\psi) + \left({p}/{\beta} + 2h\right)\lcal ''(\psi) \right)\vert\delta \psi\vert^2
 \nonumber\\
 &+& 2 {\lcal ' (\psi)}\, \delta p\, \delta \psi/{\beta}
 \bigg] .
\een
Using the relation $p(\psi)=-\lcal(\psi)$ and rearranging the terms, one can rewrite $\delta^2 F$ as follows:
\ben
\delta^2 F&=&\int dxdy \bigg[
\vert \nabla \delta \phi \vert^2 + \vert \delta v\vert^2 + \vert \nabla \delta \psi \vert^2
\nonumber\\
&+& \frac{1}{\beta}\left(\delta p - \frac{\hz\times \bp \cdot \nabla p}{B_p^2} \delta \psi\right)^2
-  \frac{\hz \times \bp}{B_p^2}\cdot \nabla \jp \, \vert \delta \psi\vert^2
\nonumber\\
&+& 2 \frac{(\hz \times \bp \cdot \nabla p)}{B_p^2} \frac{(\hz \times \bp \cdot \nabla h)}{B_p^2}\vert \delta \psi \vert^2
\bigg],
\label{2fc}
\een
where   $\bp=\nabla \psi \times \hz$ denotes the poloidal equilibrium magnetic field and  $\jp=-\lap \psi$ the equilibrium parallel current density (apart from the rescaling factor $\varepsilon$).

According to the EC method, equilibria for which $\delta^2 F $ has a definite sign, are energy stable. Analyzing the expression (\ref{2fc}), one sees that the first four terms are positive definite, whereas the last two terms can be positive or negative and, consequently,  could give rise to instability. The first of these two terms depends on the current density gradient. In the absence of magnetic field curvature effects ($h\equiv 0$), this term can be written as $-\int dxdy \vert \delta \psi \vert^2 d \jp / d \psi$. Therefore, a current profile that is monotonically decreasing in  $\psi$ has a stabilizing effect. This effect had already been identified by means of the EC method in the context of incompressible reduced MHD \cite{Haz84}.
For CRMHD, the parallel current density is no longer a magnetic flux function, because of the presence of curvature, which will then play a role in determining the stability condition.
The second term of indefinite sign, on the other hand, is stabilizing if the gradients of the equilibrium pressure  and of the magnetic curvature terms point in the same direction.

We show in the following, that the two terms of indefinite sign in (\ref{2fc}) are related to the kink and interchange instabilities for ideal MHD.

For comparison, consider the usual energy principle analysis for MHD.
Recall,  the expression for $\delta W$ in the form presented in Ref.~\cite{Gre68}:
\ben
\label{dw}
\delta W&=&\frac{1}{2}\int dxdy \bigg[
 \left( \mathbf{Q}- \frac{\bxi\cdot\nabla p}{B^2}\bB\right)^2 - \frac{\jp}{B}(\bxi \times \bB)\cdot \mathbf{Q}
 \nonumber\\
 &+& p\ (\nabla\cdot\bxi)^2+2(\bxi \cdot \nabla p)(\bxi \cdot \nabla h)
 \bigg].
\een
In (\ref{dw}) $\bxi=\bxi(\bx , t)$ indicates the displacement of a fluid element from the equilibrium position $\bx$ at the time $t$, whereas the vector $\mathbf{Q}=\rot(\bxi \times \bB)$ is the magnetic field perturbation. In writing (\ref{dw}), use has been made of the fact that the magnetic field curvature, points in the direction opposite to that of $\nabla h$.

According to the energy principle (see, e.g.,  Ref.~\cite{Gre68,Whi06}), a static MHD equilibrium, is stable if and only if $\delta W \geq 0$ for all the allowable displacements $\bxi(\bx ,t)$. It is then clear from (\ref{dw}), that only two of its terms, the second and the fourth, can have a destabilizing effect.   The second term  is associated with kink instabilities, and represents  the interaction of the parallel current with the magnetic perturbation. The fourth term, on the other hand, accounts for the interchange instability, and is destabilizing in the presence of unfavorable curvature, that is,  when $\nabla p$ and $\nabla h$ are antiparallel. A stabilizing effect, on the other hand, is due to the work needed to bend field lines, which is represented by the first term of (\ref{dw}), the energy in the field perpendicular to the equilibrium magnetic field. The parallel component of the first term of (\ref{dw}), as explained in Ref.~\cite{Gre68}, also contains a stabilizing contribution due to plasma compressibility. Finally, a stabilizing contribution due to compressibility of sound waves, comes from the third term in (\ref{dw}).

In order to compare  (\ref{dw}) and (\ref{2fc}), we must of course relate the expression (\ref{dw}), which is valid for general compressible ideal MHD, to the specific geometry and ordering of CRMHD. Using the relation $Q_{\parallel}=\bQ\cdot\bB /B=B(2 \bxi\cdot \nabla h)+(1/B)\bxi\cdot\nabla p$, obtained from Ref.~\cite{Gre68}, in the limit $\nabla\cdot \bxi=0$ (compressibility effects will be treated separately), we obtain
\ben
\left( \mathbf{Q}- \frac{\bxi\cdot\nabla p}{B^2}\bB\right)^2&=&(\bqp+2\ \bxi\cdot\nabla h\ \bB)^2
 \nonumber\\
&\simeq&
 Q_{\perp}^2+4(\bxi\cdot\nabla h)^2,  \label{qp1}
\een
where $\bqp=\mathbf{Q}-Q_{\parallel}\bB/B$ is the perturbation perpendicular to the equilibrium magnetic field and,  in the last step, we assumed that to  leading order $\bB=\hz$. The perpendicular magnetic perturbation, on the other hand, can be related to a  perturbation of the magnetic flux function $\delta \psi$ in the following way:
\beq \label{relqpsi}
\bqp=\rot(\bxi \times \bB)=\rot(\delta \psi \hz).
\eeq
Thus,  expression (\ref{qp1}) can  be rewritten as
\beq
\left( \mathbf{Q}- \frac{\bxi\cdot\nabla p}{B^2}\bB\right)^2\simeq \vert \nabla \delta \psi \vert^2+4(\bxi\cdot\nabla h)^2.  \label{qp2}
\eeq
Neglecting toroidal displacement, so that $\bxi\cdot\hz=0$, from (\ref{relqpsi}) one gets also the relation
\beq \label{psi1}
\delta \psi\,  \hz + \nabla \Upsilon= -\bxi\cdot\nabla \psi\, \hz + \bxi\times \hz,
\eeq
where $\Upsilon$ is a scalar function and where, again,  we considered the toroidal field to be constant at  leading order. The $z$ component of (\ref{psi1}) then leads to
\beq \label{reldpsi}
\delta \psi=-\bxi\cdot\nabla \psi,
\eeq
which relates an arbitrary displacement $\bxi$ to a  magnetic flux perturbation $\delta \psi$, in the presence of a given equilibrium flux function $\psi$.

Using the same approximations, the second term of  (\ref{dw}) can be written as
\ben
-\int dxdy \ \frac{\jp}{B}(\bxi \!\!&\times&\! \!\bB)\cdot \mathbf{Q} \simeq
 -\int dxdy\ \jp (\bxi \times \bB)\cdot \mathbf{Q} \nonumber\\
&=&-\int dxdy\ \jp (\bxi \times \bB)\cdot \bqp \nonumber \\
&=&-\int dxdy\ \jp (\bxi \times \bB)\cdot (\nabla\delta \psi \times \hz) \nonumber \\
&=&\int dxdy\ \delta \psi\ \bxi\cdot \nabla \jp,
 \label{jqb}
\een
where, for the last step, we integrated by parts assuming $\nabla\cdot\bxi=0$.

From Eq.~(\ref{gs}), restricted to the case $v(\psi)\equiv0$, we obtain
\beq \label{gs1}
\jp=-\fcal ' (\psi) + p'(\psi) \big({p (\psi) }/{\beta}+ 2h\big),
\eeq
which, when inserted into (\ref{jqb}) using (\ref{reldpsi}), yields
\ben \label{jql2}
&& -\int dxdy\ \frac{\jp}{B}(\bxi \times \bB)\cdot \mathbf{Q}
 \nonumber\\
&\simeq& \int dxdy\ \bigg[ \vert \delta \psi \vert^2\left(\fcal '' - p''{p}/{\beta}-{{p'}^2}/{\beta}
 -2 p''  h\right)
 \nonumber\\
&{\ }& \qquad
-2\bxi\cdot\nabla p\ \bxi \cdot\nabla h\bigg].
\een
From (\ref{gs1}), on the other hand, one can derive the relation
\ben
&&\fcal ''(\psi)- p ''(\psi){p(\psi)}/{\beta}-{{p'}^2(\psi)}/{\beta}-2 p ''(\psi) h
\nonumber\\
&=&-\frac{\hz \times \bp}{B_p^2}\cdot \nabla \jp+
2 \frac{(\hz \times \bp \cdot \nabla p)}{B_p^2} \frac{(\hz \times \bp \cdot \nabla h)}{B_p^2},
\nonumber
\een
which, when combined with (\ref{jql2}) yields
\ben
&&
-\int dxdy\ \frac{\jp}{B}(\bxi \times \bB)\cdot \mathbf{Q}
\nonumber\\
 &&  \qquad \simeq \int dxdy \bigg[ \vert \delta \psi \vert^2\bigg(-\frac{\hz \times \bp}{B_p^2}\cdot \nabla \jp
\label{jql3}
 \\
 && +  2 \frac{(\hz \times \bp \cdot \nabla p)}{B_p^2} \frac{(\hz \times \bp \cdot \nabla h)}{B_p^2} \bigg)
 -2 \bxi\cdot\nabla p\ \bxi \cdot\nabla h\bigg].
 \nonumber
\een
Finally, by combining Eqs.~(\ref{dw}), (\ref{qp2}),  and (\ref{jql3}), assuming $\nabla\cdot\bxi=0$, we obtain
\ben
\label{dw2}
\delta W\!&\simeq&\! \!\int \!dxdy \bigg(
\vert \nabla \delta \psi \vert^2+4(\bxi\cdot\nabla h)^2
-\frac{\hz \times \bp}{B_p^2}\cdot \nabla \jp  \vert\delta \psi \vert^2\nonumber \\
&&  + \ 2 \frac{(\hz \times \bp \cdot \nabla p)}{B_p^2}
 \frac{(\hz \times \bp \cdot \nabla h)}{B_p^2} \vert\delta \psi \vert^2 \bigg).
\een
By comparing (\ref{dw2}) with (\ref{2fc}), we observe that the expression for $\delta W$ contains the same destabilizing terms obtained from the EC method. Stability conditions obtained from the energy principle, are then intimately related to those obtained from the general EC method. This similarity confirms then that, for the equilibria under consideration, an indefiniteness in the sign of $\delta^2 F$ for CRMHD, can be attributed to the presence of kink or interchange instabilities.

Of course, CRMHD does not coincide exactly with the ideal MHD system to which the energy principle was applied, so that an exact one-to-one correspondence between $\delta W$ for ideal MHD and $\delta^2 F$ for CRMHD should not  be expected. However, further correspondence  can  be found. First, we observe that the stabilizing field line bending term $\vert \nabla \delta \psi \vert^2$ appears in both expression. The expression for $\delta W$ contains the additional term proportional to $(\bxi\cdot \nabla h)^2$, which is however, always positive, and therefore, not a possible source for instabilities. Second, we remark that for ideal MHD,  pressure perturbations are related to the displacement by
\beq
\delta p = -\bxi\cdot\nabla p -p\ \nabla\cdot \bxi.
\eeq
Using (\ref{reldpsi}) and the fact that $p$ is a flux function, we obtain
\ben
- p\,  \nabla\cdot \bxi &=& \delta p - p'(\psi) \delta \psi  \nonumber\\
&=&\delta p - \frac{\hz\times \bp \cdot \nabla p}{B_p^2} \delta \psi.
\een
Consequently, the fourth term under the integral in (\ref{2fc}) can be written as
\ben
\frac{1}{\beta}\left(\delta p - \frac{\hz\times \bp \cdot \nabla p}{B_p^2} \delta \psi\right)^2=\frac{p^2}{\beta} (\nabla\cdot\bxi)^2.
\een
This term is clearly reminiscent of the third term in (\ref{dw}) (recall that  $\beta$ in the denominator of $\delta^2 F$ is proportional to the background plasma pressure). This analogy then indicates that the stabilizing role of the fourth term in $\delta^2F$ for CRMHD, is due to the plasma compressibility.

Finally, we observe that the terms $\vert \nabla \delta \phi\vert^2$ and $\vert \delta v\vert^2$ in $\delta ^2 F$ obviously have no counterpart in $\delta W$. Indeed, such terms refer to variation of the kinetic energy, and are not part of the potential energy $\delta W$.

Now, we extend  the EC stability analysis for CRMHD to the more general case in which the equilibrium possesses also a toroidal flow. Therefore we impose $\gcal\equiv0$,  but  $\ncal\not\equiv 0$, so that at equilibrium,  $v(\psi)=-\ncal (\psi)$.

The second variation, then reads
\ben
 \delta^2 F&=&\int dxdy \bigg[ \vert \nabla \delta \phi \vert^2 + \vert \nabla \delta \psi \vert^2  +\vert \delta v\vert^2
 +{\vert \delta p \vert^2}/{\beta}  \nonumber\\
 &+&\left( \fcal '' (\psi) + \left({p}/{\beta} + 2h\ \right)\lcal ''(\psi)  + v\ \ncal '' (\psi)\right)\vert\delta \psi\vert^2 \nonumber  \\
 &+&2  {\lcal' (\psi)}\ \delta p\ \delta \psi/\beta  + 2 \ncal ' (\psi)\ \delta v\ \delta \psi
 \bigg].
  \label{2fvt}
\een
Using the equilibrium relations of Eqs.~(\ref{ec1})--(\ref{ec4}) with  $\gcal\equiv 0$, the expression (\ref{2fvt}) can be rewritten as
\ben
\delta^2 F&=&\int dxdy \bigg[ \vert \nabla \delta \phi \vert^2  + \vert \nabla \delta \psi \vert^2\nonumber \\
&+&  \frac{1}{\beta}\left(\delta p - \frac{\hz\times \bp \cdot \nabla p}{B_p^2} \delta \psi\right)^2
\nonumber\\
&+&  \left(\delta v - \frac{\hz\times \bp \cdot \nabla v}{B_p^2} \delta \psi\right)^2
-  \frac{\hz \times \bp}{B_p^2}\cdot \nabla \jp \vert \delta \psi\vert^2 \nonumber\\
&+& 2\,  \frac{(\hz \times \bp \cdot \nabla p)}{B_p^2} \frac{(\hz \times \bp \cdot \nabla h)}{B_p^2}\vert \delta \psi \vert^2
\bigg].  \label{2fcvt}
\een
Comparing Eq.~(\ref{2fcvt}) with (\ref{2fc}), we observe that, after the inclusion of toroidal flow, the only terms with indefinite sign are still those associated with kink and interchange instabilities. Therefore, we  conclude that, according to CRMHD, the possible instabilities for equilibria with toroidal flow obtained by extremizing the free energy functional, can still only be  of kink or interchange type.

%%%%%%%%%%%%%%%%%%%%%%%%%%%%
%%%%%%%%%%%%%%%%%%%%%%%%%%%%
\section{Analogy with the MRI model} \label{sec:analogy}

Now we consider the model for MRI in the shearing sheet approximation derived in Ref.~\cite{Jul06}. This model considers the local behavior of a conducting incompressible fluid rotating about the $z$ axis of a Cartesian coordinate system $(x,y,z)$, with $x$ and $y$ indicating the radial and azimuthal directions, respectively. The system is also supposed to be translationally invariant along the azimuthal coordinate, so that all fields depend on $x$ and $z$ only. A linear shearflow,  $\mathbf{U}_0=\sigma x \hat{y}$ along the azimuthal direction, is assumed to be maintained  and accounts for the radial variation of the angular velocity. The geometry of the system is depicted in the left panel of  Fig.~\ref{fig1}.

Neglecting dissipative terms, the model equations in a dimensionless form, are given by
\ben
\hspace{-.9 cm} \frac{\partial \omega }{\partial t}&+&2\Omega [x,v]_z +[\phi,\omega ]_z -\va [x,\lap \psi]_z
\nonumber
\\
&&\qquad +\, \va [\lap \psi , \psi]_z=0,
 \label{m1}\\
 \frac{\partial \psi}{\partial t}&+& [\phi,\psi]_z-[x,\phi]_z=0,
  \label{m2} \\
 \frac{\partial b}{\partial t}&+&[\phi , b]_z+[v, \psi]_z-[x,v]_z
 +\sigma [x,\psi ]_z=0,
 \label{m3}\\
 \frac{\partial v}{\partial t}&+&[\phi , v]_z +\va [b,\psi]_z - (2\Omega + \sigma)[x,\phi]_z
 \nonumber
\\
&&  \qquad - \va\,  [x,b]_z =0
\label{m4}.
\een
In this system $\psi(x,z)$ and $\phi(x,z)$ denote a magnetic flux function and a stream function, respectively, with $\omega =\lap \phi$.  So,   the radial and vertical components of the magnetic and velocity fields are given by $B_x=-\partial \psi/\partial z$, $B_z=\partial \psi/\partial x$ and $v_x=-\partial \phi/\partial z$ , $v_z=\partial \phi/\partial x$, respectively.  (Note, in order to facilitate  comparison with CRMHD, we interchanged the meaning of $\psi$ and $\phi$ adopted in Ref.~\cite{Jul06}.)  The fields $b(x,z)$ and $v(x,z)$, on the other hand, represent the inhomogeneous azimuthal components of the magnetic and velocity fields, respectively. The constant parameters $\Omega$, $\sigma$,  and $v_A$ denote  the angular velocity, the amplitude of the shearflow and the Alfv\'en speed based on a constant background magnetic field directed along $\hz$. We let  $[f,g]_z=-\hat{y}\cdot \nabla f \times \nabla g$, which is the  canonical bracket  with respect to the variables $x$ and $z$ between two functions $f$ and $g$.

We transform to   the following new variables:
\beq  \label{trsf}
\tilde{\phi}=\phi, \quad \tilde{\psi}=v_A(\psi + x), \quad \tilde{b}=v_A b , \quad \tilde{v}=v+\sigma x,
\eeq
where $\tilde{\psi}$ is a flux function that  includes a  constant background $z$-component of the magnetic field and  $\tilde{v}$ is the $y$-component of a velocity field that includes a  linear shearflow.

In terms of the variables $(\tilde{\phi},\tilde{\psi},\tilde{b},\tilde{v})$, the system of Eqs.~(\ref{m1})--(\ref{m4}) becomes
\ben
\hspace{-.8 cm} \frac{\partial \tilde{\omega}}{\partial t}&+&[\tilde{\phi}, \tilde{\omega}]_z +[\lap \tilde{\psi} , \tilde{\psi}]_z -2\Omega [\tilde{v},x]_z =0, \label{m1v}\\
\frac{\partial \tilde{\psi}}{\partial t}&+&[\tilde{\phi},\tilde{\psi} ]_z=0,  \label{m2v} \\
\frac{\partial \tilde{b}}{\partial t}&+&[\tilde{\phi} , \tilde{b} ]_z +[\tilde{v}, \tilde{\psi}]_z=0, \label{m3v}\\
\frac{\partial \tilde{v}}{\partial t}&+&[\tilde{\phi} , \tilde{v} ]_z + [\tilde{b},\tilde{\psi}]_z- 2\Omega [x,\tilde{\phi}]_z =0 \label{m4v},
\een
where $\tilde\omega=\lap \tilde\phi$. Comparing Eqs.~(\ref{m1v})--(\ref{m4v}) with the CRMHD model of Eqs.~(\ref{c1})--(\ref{c4}), it is evident that the two models are identical provided one sets $\beta=1$ and makes the following identifications:
\ben
\tilde{\phi}\leftrightarrow \phi, \qquad \tilde{\psi}\leftrightarrow \psi, \qquad \tilde{b}\leftrightarrow v, \qquad \tilde{v}\leftrightarrow p,\\
y \leftrightarrow z, \qquad \Omega x \leftrightarrow h ,\qquad [\  ,\  ]_z \leftrightarrow [\  ,\  ]_y.
\een

It emerges then, from this comparison,  that there is an analogy between  the role played by the angular rotation in the MRI system  with  that of magnetic curvature in CRMHD.   Indeed, the last terms in Eqs.~(\ref{m1v}) and (\ref{m4v}), which are due to the Coriolis force acting on the fluid, match the last terms of Eqs.~(\ref{c1}) and (\ref{c4}), which come from the toroidal component of the Lorentz force (together with the assumption that the pressure and the diamagnetic correction of the toroidal magnetic field are proportional) and  the compressibility of the $\mathbf{E}\times\mathbf{B}$ flow in the presence of a curved magnetic field, respectively. The third term in the pressure equation (\ref{c4}), which also accounts for compressibility effects, matches the magnetic tension term in the azimuthal velocity equation (\ref{m4v}). As pointed out in Ref.~\cite{Haz85}, the inclusion of compressibility terms (i.e., those proportional to $\beta$) in CRMHD  is not strictly consistent with the ordering adopted to derive the model. However, such terms were  retained in the model due to their ``special qualitative importance'' \cite{Haz85}. It is worth noticing that, without such terms, the analogy between the pressure equation of CRMHD and the azimuthal velocity equation of the (incompressible) MRI model (which does strictly follows an imposed ordering) would not exist.  We remark also that the toroidal velocity of CRMHD, plays a role analogous to that of the azimuthal magnetic field, and in particular, the last term of Eq.~(\ref{c3}), expressing the parallel pressure gradient, mirrors the last term of Eq.~(\ref{m3v}), which represents the parallel gradient of the azimuthal velocity. The analogy between the  vorticity equation (\ref{c1}) and the  Ohm's law Eq.~(\ref{c2}) of CRMHD and their counterparts in the MRI model is more natural, apart from the already discussed curvature/Coriolis terms.

In spite of this formal analogy between the models, however,  we remark that, under the transformation Eq.~(\ref{trsf}), the new variables do not satisfy the same radial boundary conditions as the old.

%%%%%%%%%%%%%%%%%%%%%%%%%%%%
%%%%%%%%%%%%%%%%%%%%%%%%%%%%
\section{MRI model Hamiltonian structure and stability}
 \label{sec:mristab}

For the Hamiltonian structure of  the MRI model (\ref{m1})--(\ref{m4}), the boundary conditions imposed on the fields are analogous to those imposed for CRMHD;  that is,  vanishing Dirichlet boundary conditions along $x$ and periodic along $z$.  With these assumptions,  the Hamiltonian of the system is given by the functional
\ben \label{hmri}
H&=&\frac{1}{2}\!\int \!dx dz \big( \vert \nabla \phi \vert^2 + (v + \sigma x)^2 \nonumber\\
&& \qquad + \, \va \vert \nabla \psi \vert^2 + \va b^2 \big),
\een
where each term has  a clear interpretation:  the first two terms account for the kinetic energy (including that of the shearflow) and last two terms represent the magnetic energy.  Unlike CRMHD, no internal energy term is present, as is typical of incompressible models  where the pressure is not an independent dynamical variable.

The Poisson bracket is given by
\ben
\{F,G\}\!&=&\!\!\int dx dz \Big[
\omega [F_{\omega},G_{\omega}]_z+b\big([F_{\omega},G_b]_z+[F_b,G_{\omega}]_z\big) \nonumber\\
&&  +\,  \big(v+(2 \Omega + \sigma)x\big) \big([F_{\omega},G_v]_z+[F_v, G_{\omega}]_z\big)\nonumber\\
&& +\,  \big(\psi+x\big)\big([F_{\omega},G_{\psi}]_z
 \label{bmri} \\
&& \quad +\, [F_{\psi},G_{\omega}]_z-[F_b,G_v]_z-[F_v,G_b]_z)\big)\Big],
\nonumber
\een
which satisfies all the requisite Poisson bracket properties.   Together with the Hamiltonian (\ref{hmri}),  this bracket yields the MRI model (\ref{m1})--(\ref{m4}). Moreover, it possesses the same structure as  the CRMHD bracket  of Eq.~(\ref{bcrmhd}). Consequently, the Casimirs of (\ref{bmri}), which are given by the following four infinite families:
\ben \label{casmri}
C_1&=&\int dxdz\ \mathcal{F}(\psi+x),
\nonumber\\
C_2&=&\int dxdz\ b\ \mathcal{N}(\psi+x),
\nonumber \\
C_3&=&\int dxdz\ \mathcal{L}(\psi+x)\big(v+(2\Omega +\sigma)x\big),
\nonumber
\\
C_4 &=&\int dxdz \big(\omega\ \mathcal{G}(\psi+x)
\nonumber\\
&&\quad  -b\,    \mathcal{G}'(\psi+x)\left(v+(2\Omega + \sigma)x \right)\big),
\nonumber
\een
are analogous to those of Eqs.~(\ref{ccrmhd}).  We find here, again the conservation of magnetic flux, expressed by the family $C_1$,  remnants of magnetic helicity, $C_2$,  and cross-helicity conservation, $C_3$ and $C_4$.

The EC stability analysis proceeds as for CRMHD.
Extremizing the free energy functional leads to the following equilibrium equations:
\ben
\phi&=&\gcal (\psi+x), \label{em1} \\
\va \lap \psi&=&\fcal ' (\psi+x) + b\ \ncal ' (\psi+x) \nonumber\\
&& \hspace{-.2 cm} + \big(v+(2\Omega  +\sigma)x\big)\lcal ' (\psi+x) + \omega\ \gcal ' (\psi+x) \nonumber\\
&& - \, b\,  \gcal '' (\psi+x) \big((v+(2\Omega +\sigma)x\big), \label{em2} \\
\va b&=& -\ncal (\psi+x) \nonumber\\
&& +\,  \gcal ' (\psi+x) \big(v+(2\Omega +\sigma)x \big), \label{em3} \\
v+\sigma x&=&-\lcal (\psi+x) +b\ \gcal ' (\psi+x).
 \label{em4}
\een
from which a plethora of equilibrium states are possible.   However, as a  first case for this model, we consider the trivial equilibrium state
\ben  \label{baseqm}
\phi\equiv0, \qquad \psi\equiv0, \qquad b\equiv0, \qquad v\equiv0,
\een
i.e.,  only the background constant magnetic fields and the linear shearflow are present at equilibrium.

The state (\ref{baseqm}) corresponds to the following choice for the free functions of the Casimirs (\ref{casmri}):
\ben
\fcal&=&\sigma\frac{2\Omega+\sigma}{2}(\psi+x)^2,
 \qquad \ncal\equiv0,
\nonumber \\
 \qquad \lcal&=&-\sigma(\psi+x), \qquad \gcal\equiv0.
 \nonumber
\een
The second variation of the corresponding free energy functional is then given by
\ben
\delta^2 F&=&\int dxdz \Big[
\vert \nabla \delta \phi\vert^2 +\vert\delta v\vert^2 +\va \vert \nabla \delta \psi\vert^2
\nonumber\\
&&+\, \va \vert \delta b\vert^2+(\sigma^2 + 2 \Omega \sigma)\vert\delta \psi\vert^2
-2\sigma\, \delta \psi \delta v
\Big],
\nonumber
\een
which can be rearranged as
\ben
\delta^2 F&=&\int dxdz\Big[\,  \vert \nabla \delta \phi\vert^2 + \va \vert \nabla \delta \psi\vert^2
\nonumber
\\
&& \hspace{-.4 cm} +\, \va \vert \delta b\vert^2+\vert \sigma \delta \psi
- \delta v\vert^2 +2\Omega \sigma \vert\delta \psi\vert^2
\Big].
\label{2fbas}
\een
It is clear from  (\ref{2fbas})  that the only possible nonpositive-definite term, that is  the only possible source for instability, comes from the last term of (\ref{2fbas}). Indeed,   the second variation is not manifestly positive  definite if $\Omega \sigma <0$, which corresponds (assuming $\Omega >0$) to an angular velocity radially decreasing outward. This is, in fact,  the condition for MRI in the shearing sheet approximation derived in Ref.~\cite{Jul06}.  (Note, application of the Poincar\' e inequality for suitable $\delta \psi$ gives another stability condition that depends effectively on the system size.)

We remark that in CRMHD the  trivial state $\phi\equiv\psi\equiv v\equiv p \equiv 0$,  analogous to (\ref{baseqm}), corresponds to $\fcal ' \equiv \gcal\equiv \ncal \equiv \lcal\equiv 0$. From (\ref{2fc1}),  one concludes immediately that such an equilibrium is always stable.  Notice, however,  whereas the MRI model embodies the ingredients for the MRI (identified by the parameters $\Omega$ and $\sigma$, in addition, of course, to the imposed background magnetic field) in its equations of motion, this is not the case for CRMHD, which accounts for the field curvature but not necessarily for the pressure gradient, which is required for the interchange instability.  In order to allow for this instability to take place, it is therefore necessary to have a nontrivial solution for the pressure in the equilibrium equations (\ref{ec1})--(\ref{ec4}). However, apart  from the trivial states,  the analogy between the MRI and the interchange instability, already suggested in Sec.~\ref{sec:analogy}, becomes evident if one considers more general equilibrium states. Indeed, by taking advantage of  the analogies between the Hamiltonian and the Casimirs in the two models, we can readily write the second variation for the free energy functional $F=H+C_1+C_2+C_3$ for  the case of the MRI model. The resulting expression is
\ben
\delta^2 F&=&\int dxdz \Big[ \vert \nabla \delta \phi \vert^2
+ \va\vert \nabla \delta \psi \vert^2   \label{2fmvt} \\
&&\hspace{-.7 cm}
+\,  \Big(\delta v - \frac{\hy\times \tbp \cdot \nabla (v+\sigma x)}{\tilde{B}_p^2} \delta \psi\Big)^2
\nonumber\\
&&\hspace{-.7 cm}
 +\,  \va\Big(\delta b - \frac{\hy\times \tbp \cdot \nabla b}{\tilde{B}_p^2} \delta \psi\Big)^2
 -\frac{\hy \times \tbp}{\tilde{B}_p^2}\cdot \nabla \tilde{j}_{\parallel} \vert \delta \psi\vert^2
\nonumber\\
&&\hspace{-.7 cm}
+ 2\,  \frac{(\hy \times \tbp \cdot \nabla (v+\sigma x))}{\tilde{B}_p^2}
 \frac{(\hy \times \tbp \cdot \nabla (2\Omega x))}{\tilde{B}_p^2}\vert \delta \psi \vert^2 \Big],
 \nonumber
\een
where $\tbp=\nabla(\psi+x)\times\hy$ and $\tilde{j}_{\parallel}=-(\partial^2 \psi/\partial x^2+\partial^2 \psi/\partial z^2)$.
It is evident upon comparing (\ref{2fmvt}) with (\ref{2fcvt}), that the term responsible for the interchange instability in CRMHD is analogous to the last term of (\ref{2fmvt}), which is responsible for the MRI. If the equilibrium azimuthal flow is reduced only to the imposed shearflow (i.e., $v\equiv0$), then the latter term yields the above mentioned instability condition $\Omega \sigma <0$. However, Eq.~(\ref{2fmvt}) offers now a stability condition that applies to more general equilibrium flows.  Also, the next to  last term in Eq.~(\ref{2fmvt}) indicates the possibility for a kink-type instability that the rotating magnetized disk can undergo in the presence of an equilibrium current density.

Finally, note that if the presence of the magnetic field is removed from the MRI model, then the system of Eqs.~(\ref{m1})--(\ref{m4}) reduces the ideal  fluid model
\ben
\frac{\partial \omega}{\partial t}&+&2\Omega[x,v]_z+[\phi,\omega]_z=0, \label{f1}\\
\frac{\partial v}{\partial t}&+&[\phi , v]_z- (2\Omega + \sigma)[x,\phi]_z=0 \label{f2}.
\een
This model is characterized by the Hamiltonian
\beq
H=\frac{1}{2}\int dxdz (\vert\nabla\phi\vert^2+(v+\sigma x)^2),
\eeq
and the bracket
\ben
\{F,G\}&=&\int dxdz \Big[\omega [F_{\omega},G_{\omega}]_z
\nonumber
\\
&&
+\, \big(v+(2 \Omega + \sigma)x\big)\big([F_{\omega},G_v]_z+[F_v, G_{\omega}]_z\big)\Big],
\nonumber
\een
which possesses the two families of Casimirs
\ben
C_1&=&\int dxdz\ \fcal\big(v+(2\Omega+\sigma)x\big),
\nonumber
\\
 C_2&=&\int dx dz\ \omega\ \gcal\big(v+(2\Omega + \sigma)x\big),
 \nonumber
\een
for arbitrary $\fcal$ and $\gcal$.
The second variation of  the free energy functional, for this model, evaluated at the trivial equilibrium $\phi\equiv v\equiv0$ (corresponding to  $\fcal(v+(2\Omega+\sigma)x)=-(\sigma/(4\Omega+2\sigma))(v+(2\Omega+\sigma)x)^2$ and $\gcal\equiv0$), reads
\ben
\delta^2 F=\int dxdz \left(\vert \nabla\delta \phi\vert^2+\frac{2\Omega}{2\Omega+\sigma}\vert\delta v\vert^2\right).
\een
One can then easily see that, if one assumes, as before, $\Omega>0$, energy stability is achieved  if
\ben
2\Omega+\sigma>0,
\een
which is namely the Rayleigh hydrodynamic stability condition for the rotating fluid that one recovers from Ref.~\cite{Jul10}. The regime of relevance for MRI is of course that for which the equilibrium is hydrodynamically stable (i.e.,  $2\Omega+\sigma>0$), but the presence of the imposed vertical magnetic field makes it energy unstable, which occurs when $\sigma<0$, as shown above.

%%%%%%%%%%%%%%%%%%%%%%%%%%%%
%%%%%%%%%%%%%%%%%%%%%%%%%%%%
\section{Conclusions} \label{sec:concl}

We have presented analytical results concerning the Hamiltonian structure and the stability properties of two reduced models, one of interest for tokamak dynamics and the other for accretion disks dynamics. Our analysis revealed analogies   between the two models and, in particular,  analogous instability processes that they can describe.  We also demonstrated   a   connection between the standard MHD energy principle and the EC stability method.

More specifically, after reviewing the Hamiltonian structure of the CRMHD model, we derived energy stability conditions for a wide class of equilibria obtained from a variational principle. Two terms of $\delta^2 F$ that can lead to instability were identified. The physical meaning of these terms was exposed by tracing their meaning back to the  ideal MHD $\delta W$ analysis, traditionally adopted in plasma physics.   Thus, these terms were seen to  corresponding to the kink and interchange instabilities.  Similarly, in the expression for $\delta^2 F$, the stabilizing roles of  compressibility and  magnetic field line bending were noted.  By comparing the cases of equilibria with and without toroidal flow, we concluded  that instabilities (such as, e.g.,  Kelvin-Helmholtz instabilities) introduced by the addition of  sheared toroidal flow can be associated with the terms responsible for the kink and interchange mechanisms. Thus, instabilities instigated  by toroidal flow can then still be detected by considering the direction of current and density gradients, with respect to magnetic flux gradients and magnetic curvature, respectively.

Next, we described the reduced model of  Ref.~\cite{Jul06}, introduced  for investigating MRI in the shearing sheet approximation. We provided a formal mapping between CRMHD and this  MRI model. The roles played in the latter by the azimuthal magnetic and velocity components, turn out to correspond to those played by toroidal velocity and pressure, respectively, in CRMHD.  We remarked that, curiously, the analogy between these  two models is only possible  because the  compressibility term has been kept {\it ad hoc}  in CRMHD, although  this term is  negligible according to the imposed ordering.

By taking advantage of the formal analogy between the models, we obtained the Hamiltonian structure for the MRI model  and identified  four infinite families of Casimir invariants expressing conservation laws.

We then applied the EC method to the MRI model, first for a trivial equilibrium  state usually adopted for MRI studies.  For this equilibrium we  recovered the instability condition $\Omega \sigma <0$,  derived in Ref.~\cite{Jul06}, that  corresponds to the usual condition of angular velocity decreasing radially outward.   We then extended the analysis to much more general equilibria.  From the analogy $\delta^2 F$ was immediate; whence we concluded that the destabilizing (nonpositive-definite) term responsible for MRI is analogous to the term responsible for interchange instability in CRMHD.   Thus,  the role played by the magnetic curvature in tokamaks is analogous to that played by  the gradient of the angular velocity in accretion disks. This term was seen to give  rise to interchange or magnetorotational instability, respectively, when   orientation is  in an unfavorable direction, with respect to the pressure or azimuthal velocity field gradients. The analogy was extended further, indicating  that  accretion disk equilibria can undergo instability  analogous of the kink instability, when the gradient of the parallel current density points toward an unfavorable direction with respect to $\nabla \psi$.

More generally, the analogy permits the  transference of   knowledge about  CRMHD stability results  to the accretion disk model, and vice versa. Thus the  Hamiltonian structure of the models provides a convenient and unifying framework for investigating properties such as conservation laws and stability conditions for both models.

\acknowledgments
This work was supported by the European Community under the contracts of Association between EURATOM, CEA, and the French Research Federation for fusion studies. The views and opinions expressed herein do not necessarily reflect those of the European Commission. E.T. received financial support from the Agence Nationale de la Recherche (ANR GYPSI). P.J.M.  was supported  by the US Department of Energy Contract No.~DE-FG03-96ER-54346 and acknowledges the warm hospitality of the nonlinear dynamics group of the Centre de Physique Th\'eorique  of the CNRS facility in Luminy.

\bibliographystyle{apsrev4-1}
%\bibliography{MyBiblioMRI_new}

\begin{thebibliography}{37}%
\makeatletter
\providecommand \@ifxundefined [1]{%
 \@ifx{#1\undefined}
}%
\providecommand \@ifnum [1]{%
 \ifnum #1\expandafter \@firstoftwo
 \else \expandafter \@secondoftwo
 \fi
}%
\providecommand \@ifx [1]{%
 \ifx #1\expandafter \@firstoftwo
 \else \expandafter \@secondoftwo
 \fi
}%
\providecommand \natexlab [1]{#1}%
\providecommand \enquote  [1]{``#1''}%
\providecommand \bibnamefont  [1]{#1}%
\providecommand \bibfnamefont [1]{#1}%
\providecommand \citenamefont [1]{#1}%
\providecommand \href@noop [0]{\@secondoftwo}%
\providecommand \href [0]{\begingroup \@sanitize@url \@href}%
\providecommand \@href[1]{\@@startlink{#1}\@@href}%
\providecommand \@@href[1]{\endgroup#1\@@endlink}%
\providecommand \@sanitize@url [0]{\catcode `\\12\catcode `\$12\catcode
  `\&12\catcode `\#12\catcode `\^12\catcode `\_12\catcode `\%12\relax}%
\providecommand \@@startlink[1]{}%
\providecommand \@@endlink[0]{}%
\providecommand \url  [0]{\begingroup\@sanitize@url \@url }%
\providecommand \@url [1]{\endgroup\@href {#1}{\urlprefix }}%
\providecommand \urlprefix  [0]{URL }%
\providecommand \Eprint [0]{\href }%
\providecommand \doibase [0]{http://dx.doi.org/}%
\providecommand \selectlanguage [0]{\@gobble}%
\providecommand \bibinfo  [0]{\@secondoftwo}%
\providecommand \bibfield  [0]{\@secondoftwo}%
\providecommand \translation [1]{[#1]}%
\providecommand \BibitemOpen [0]{}%
\providecommand \bibitemStop [0]{}%
\providecommand \bibitemNoStop [0]{.\EOS\space}%
\providecommand \EOS [0]{\spacefactor3000\relax}%
\providecommand \BibitemShut  [1]{\csname bibitem#1\endcsname}%
\let\auto@bib@innerbib\@empty
%</preamble>
\bibitem [{\citenamefont {Bernstein}\ \emph {et~al.}(1958)\citenamefont
  {Bernstein}, \citenamefont {Frieman}, \citenamefont {Kruskal},\ and\
  \citenamefont {Kulsrud}}]{Ber58}%
  \BibitemOpen
  \bibfield  {author} {\bibinfo {author} {\bibfnamefont {I.~B.}\ \bibnamefont
  {Bernstein}}, \bibinfo {author} {\bibfnamefont {E.~A.}\ \bibnamefont
  {Frieman}}, \bibinfo {author} {\bibfnamefont {M.~D.}\ \bibnamefont
  {Kruskal}}, \ and\ \bibinfo {author} {\bibfnamefont {R.~M.}\ \bibnamefont
  {Kulsrud}},\ }\href@noop {} {\bibfield  {journal} {\bibinfo  {journal}
  {Proceedings of Royal Society London A}\ }\textbf {\bibinfo {volume} {244}},\
  \bibinfo {pages} {17} (\bibinfo {year} {1958})}\BibitemShut {NoStop}%
\bibitem [{\citenamefont {Freidberg}(1987)}]{Fre87}%
  \BibitemOpen
  \bibfield  {author} {\bibinfo {author} {\bibfnamefont {J.~P.}\ \bibnamefont
  {Freidberg}},\ }\href@noop {} {\emph {\bibinfo {title} {{Ideal
  Magnetohydrodynamics }}}}\ (\bibinfo  {publisher} {Plenum Press, New York},\
  \bibinfo {year} {1987})\BibitemShut {NoStop}%
\bibitem [{\citenamefont {Kruskal}\ and\ \citenamefont
  {Oberman}(1958)}]{Kru58}%
  \BibitemOpen
  \bibfield  {author} {\bibinfo {author} {\bibfnamefont {M.~D.}\ \bibnamefont
  {Kruskal}}\ and\ \bibinfo {author} {\bibfnamefont {C.}~\bibnamefont
  {Oberman}},\ }\href@noop {} {\bibfield  {journal} {\bibinfo  {journal} {Phys.
  Fluids}\ }\textbf {\bibinfo {volume} {1}},\ \bibinfo {pages} {275} (\bibinfo
  {year} {1958})}\BibitemShut {NoStop}%
\bibitem [{\citenamefont {Morrison}\ and\ \citenamefont
  {Greene}(1980)}]{Mor80}%
  \BibitemOpen
  \bibfield  {author} {\bibinfo {author} {\bibfnamefont {P.~J.}\ \bibnamefont
  {Morrison}}\ and\ \bibinfo {author} {\bibfnamefont {J.~M.}\ \bibnamefont
  {Greene}},\ }\href@noop {} {\bibfield  {journal} {\bibinfo  {journal} {Phys.
  Rev. Lett.}\ }\textbf {\bibinfo {volume} {45}},\ \bibinfo {pages} {790}
  (\bibinfo {year} {1980})}\BibitemShut {NoStop}%
\bibitem [{\citenamefont {Morrison}\ and\ \citenamefont
  {Hazeltine}(1984{\natexlab{a}})}]{Mor84}%
  \BibitemOpen
  \bibfield  {author} {\bibinfo {author} {\bibfnamefont {P.~J.}\ \bibnamefont
  {Morrison}}\ and\ \bibinfo {author} {\bibfnamefont {R.~D.}\ \bibnamefont
  {Hazeltine}},\ }\href@noop {} {\bibfield  {journal} {\bibinfo  {journal}
  {Phys. Fluids}\ }\textbf {\bibinfo {volume} {27}},\ \bibinfo {pages} {886}
  (\bibinfo {year} {1984}{\natexlab{a}})}\BibitemShut {NoStop}%
\bibitem [{\citenamefont {Marsden}\ and\ \citenamefont
  {Morrison}(1984)}]{Mar84}%
  \BibitemOpen
  \bibfield  {author} {\bibinfo {author} {\bibfnamefont {J.~E.}\ \bibnamefont
  {Marsden}}\ and\ \bibinfo {author} {\bibfnamefont {P.~J.}\ \bibnamefont
  {Morrison}},\ }\href@noop {} {\bibfield  {journal} {\bibinfo  {journal}
  {Contemp. Math.}\ }\textbf {\bibinfo {volume} {28}},\ \bibinfo {pages} {133}
  (\bibinfo {year} {1984})}\BibitemShut {NoStop}%
\bibitem [{\citenamefont {Weinstein}(1983)}]{Wei83}%
  \BibitemOpen
  \bibfield  {author} {\bibinfo {author} {\bibfnamefont {A.}~\bibnamefont
  {Weinstein}},\ }\href@noop {} {\bibfield  {journal} {\bibinfo  {journal}
  {Phys. Fluids}\ }\textbf {\bibinfo {volume} {26}},\ \bibinfo {pages} {388}
  (\bibinfo {year} {1983})}\BibitemShut {NoStop}%
\bibitem [{\citenamefont {Hazeltine}\ \emph {et~al.}(1987)\citenamefont
  {Hazeltine}, \citenamefont {Hsu},\ and\ \citenamefont {Morrison}}]{Haz87}%
  \BibitemOpen
  \bibfield  {author} {\bibinfo {author} {\bibfnamefont {R.~D.}\ \bibnamefont
  {Hazeltine}}, \bibinfo {author} {\bibfnamefont {C.~T.}\ \bibnamefont {Hsu}},
  \ and\ \bibinfo {author} {\bibfnamefont {P.~J.}\ \bibnamefont {Morrison}},\
  }\href@noop {} {\bibfield  {journal} {\bibinfo  {journal} {Physics of
  Fluids}\ }\textbf {\bibinfo {volume} {30}},\ \bibinfo {pages} {3204}
  (\bibinfo {year} {1987})}\BibitemShut {NoStop}%
\bibitem [{\citenamefont {Schep}\ \emph {et~al.}(1994)\citenamefont {Schep},
  \citenamefont {Pegoraro},\ and\ \citenamefont {Kuvshinov}}]{Sch94}%
  \BibitemOpen
  \bibfield  {author} {\bibinfo {author} {\bibfnamefont {T.}~\bibnamefont
  {Schep}}, \bibinfo {author} {\bibfnamefont {F.}~\bibnamefont {Pegoraro}}, \
  and\ \bibinfo {author} {\bibfnamefont {B.}~\bibnamefont {Kuvshinov}},\
  }\href@noop {} {\bibfield  {journal} {\bibinfo  {journal} {Phys. Plasmas}\
  }\textbf {\bibinfo {volume} {1}},\ \bibinfo {pages} {2843} (\bibinfo {year}
  {1994})}\BibitemShut {NoStop}%
\bibitem [{\citenamefont {Tassi}\ \emph {et~al.}(2008)\citenamefont {Tassi},
  \citenamefont {Morrison}, \citenamefont {Waelbroeck},\ and\ \citenamefont
  {Grasso}}]{Tas08}%
  \BibitemOpen
  \bibfield  {author} {\bibinfo {author} {\bibfnamefont {E.}~\bibnamefont
  {Tassi}}, \bibinfo {author} {\bibfnamefont {P.~J.}\ \bibnamefont {Morrison}},
  \bibinfo {author} {\bibfnamefont {F.~L.}\ \bibnamefont {Waelbroeck}}, \ and\
  \bibinfo {author} {\bibfnamefont {D.}~\bibnamefont {Grasso}},\ }\href@noop {}
  {\bibfield  {journal} {\bibinfo  {journal} {Plasma Phys. and Contr. Fus.}\
  }\textbf {\bibinfo {volume} {36}},\ \bibinfo {pages} {085014} (\bibinfo
  {year} {2008})}\BibitemShut {NoStop}%
\bibitem [{\citenamefont {Tronci}(2010)}]{Tro10}%
  \BibitemOpen
  \bibfield  {author} {\bibinfo {author} {\bibfnamefont {C.}~\bibnamefont
  {Tronci}},\ }\href@noop {} {\bibfield  {journal} {\bibinfo  {journal} {J.
  Phys. A: Math. Theor.}\ }\textbf {\bibinfo {volume} {43}},\ \bibinfo {pages}
  {375501} (\bibinfo {year} {2010})}\BibitemShut {NoStop}%
\bibitem [{\citenamefont {Morrison}(1982)}]{Mor82a}%
  \BibitemOpen
  \bibfield  {author} {\bibinfo {author} {\bibfnamefont {P.~J.}\ \bibnamefont
  {Morrison}},\ }\href@noop {} {\bibfield  {journal} {\bibinfo  {journal} {AIP
  Conf. Proc.}\ }\textbf {\bibinfo {volume} {88}},\ \bibinfo {pages} {13}
  (\bibinfo {year} {1982})},\ \bibinfo {note}
  {doi:http://dx.doi.org/10.1063/1.33633}\BibitemShut {NoStop}%
\bibitem [{\citenamefont {Morrison}(1998)}]{Mor98}%
  \BibitemOpen
  \bibfield  {author} {\bibinfo {author} {\bibfnamefont {P.~J.}\ \bibnamefont
  {Morrison}},\ }\href@noop {} {\bibfield  {journal} {\bibinfo  {journal} {Rev.
  Mod. Phys.}\ }\textbf {\bibinfo {volume} {70}},\ \bibinfo {pages} {467}
  (\bibinfo {year} {1998})}\BibitemShut {NoStop}%
\bibitem [{\citenamefont {Morrison}(2005)}]{Mor05}%
  \BibitemOpen
  \bibfield  {author} {\bibinfo {author} {\bibfnamefont {P.~J.}\ \bibnamefont
  {Morrison}},\ }\href@noop {} {\bibfield  {journal} {\bibinfo  {journal}
  {Phys. Plasmas}\ }\textbf {\bibinfo {volume} {12}},\ \bibinfo {pages}
  {058102} (\bibinfo {year} {2005})}\BibitemShut {NoStop}%
\bibitem [{\citenamefont {Hazeltine}\ \emph {et~al.}(1984)\citenamefont
  {Hazeltine}, \citenamefont {Holm}, \citenamefont {Marsden},\ and\
  \citenamefont {Morrison}}]{Haz84}%
  \BibitemOpen
  \bibfield  {author} {\bibinfo {author} {\bibfnamefont {R.~D.}\ \bibnamefont
  {Hazeltine}}, \bibinfo {author} {\bibfnamefont {D.~D.}\ \bibnamefont {Holm}},
  \bibinfo {author} {\bibfnamefont {J.~E.}\ \bibnamefont {Marsden}}, \ and\
  \bibinfo {author} {\bibfnamefont {P.~J.}\ \bibnamefont {Morrison}},\
  }\href@noop {} {\bibfield  {journal} {\bibinfo  {journal} {Proceedings of
  International Conference on Plasma Physics (ICPP) Lausanne}\ }\textbf
  {\bibinfo {volume} {1}},\ \bibinfo {pages} {203} (\bibinfo {year}
  {1984})}\BibitemShut {NoStop}%
\bibitem [{\citenamefont {Holm}\ \emph {et~al.}(1985)\citenamefont {Holm},
  \citenamefont {Marsden}, \citenamefont {Ratiu},\ and\ \citenamefont
  {Weinstein}}]{Hol85}%
  \BibitemOpen
  \bibfield  {author} {\bibinfo {author} {\bibfnamefont {D.~D.}\ \bibnamefont
  {Holm}}, \bibinfo {author} {\bibfnamefont {J.~E.}\ \bibnamefont {Marsden}},
  \bibinfo {author} {\bibfnamefont {T.~S.}\ \bibnamefont {Ratiu}}, \ and\
  \bibinfo {author} {\bibfnamefont {A.}~\bibnamefont {Weinstein}},\ }\href@noop
  {} {\bibfield  {journal} {\bibinfo  {journal} {Physics Reports}\ }\textbf
  {\bibinfo {volume} {123}},\ \bibinfo {pages} {2} (\bibinfo {year}
  {1985})}\BibitemShut {NoStop}%
\bibitem [{\citenamefont {Morrison}\ and\ \citenamefont
  {Eliezer}(1986)}]{Mor86}%
  \BibitemOpen
  \bibfield  {author} {\bibinfo {author} {\bibfnamefont {P.~J.}\ \bibnamefont
  {Morrison}}\ and\ \bibinfo {author} {\bibfnamefont {S.}~\bibnamefont
  {Eliezer}},\ }\href@noop {} {\bibfield  {journal} {\bibinfo  {journal} {Phys.
  Rev. A}\ }\textbf {\bibinfo {volume} {33}},\ \bibinfo {pages} {4205}
  (\bibinfo {year} {1986})}\BibitemShut {NoStop}%
\bibitem [{\citenamefont {Julien}\ and\ \citenamefont
  {Knobloch}(2006)}]{Jul06}%
  \BibitemOpen
  \bibfield  {author} {\bibinfo {author} {\bibfnamefont {K.}~\bibnamefont
  {Julien}}\ and\ \bibinfo {author} {\bibfnamefont {E.}~\bibnamefont
  {Knobloch}},\ }\href@noop {} {\bibfield  {journal} {\bibinfo  {journal} {EAS
  Pub. Series}\ }\textbf {\bibinfo {volume} {21}},\ \bibinfo {pages} {81}
  (\bibinfo {year} {2006})}\BibitemShut {NoStop}%
\bibitem [{\citenamefont {Morrison}\ and\ \citenamefont
  {Hazeltine}(1984{\natexlab{b}})}]{Mor84b}%
  \BibitemOpen
  \bibfield  {author} {\bibinfo {author} {\bibfnamefont {P.}~\bibnamefont
  {Morrison}}\ and\ \bibinfo {author} {\bibfnamefont {R.~D.}\ \bibnamefont
  {Hazeltine}},\ }\href@noop {} {\bibfield  {journal} {\bibinfo  {journal}
  {Proceedings of the Sherwood Theory Conference (Lawrence Livermore National
  Laboratory, Incline Village, Nevada)}\ }\textbf {\bibinfo {volume} {R12}},\
  \bibinfo {pages} {2} (\bibinfo {year} {1984}{\natexlab{b}})}\BibitemShut
  {NoStop}%
\bibitem [{\citenamefont {Velikhov}(36)}]{Vel59}%
  \BibitemOpen
  \bibfield  {author} {\bibinfo {author} {\bibfnamefont {E.~P.}\ \bibnamefont
  {Velikhov}},\ }\href@noop {} {\bibfield  {journal} {\bibinfo  {journal} {Sov.
  Phys. JETP}\ }\textbf {\bibinfo {volume} {16}},\ \bibinfo {pages} {1398}
  (\bibinfo {year} {36})}\BibitemShut {NoStop}%
\bibitem [{\citenamefont {Chandrasekhar}(1960)}]{Cha60}%
  \BibitemOpen
  \bibfield  {author} {\bibinfo {author} {\bibfnamefont {S.}~\bibnamefont
  {Chandrasekhar}},\ }\href@noop {} {\bibfield  {journal} {\bibinfo  {journal}
  {Proceedings of Natl Acad. Sci. USA}\ }\textbf {\bibinfo {volume} {46}},\
  \bibinfo {pages} {253} (\bibinfo {year} {1960})}\BibitemShut {NoStop}%
\bibitem [{\citenamefont {Acheson}(1973)}]{Ach73fl}%
  \BibitemOpen
  \bibfield  {author} {\bibinfo {author} {\bibfnamefont {D.~J.}\ \bibnamefont
  {Acheson}},\ }\href@noop {} {\bibfield  {journal} {\bibinfo  {journal} {J.
  Fluid Mech.}\ }\textbf {\bibinfo {volume} {61}},\ \bibinfo {pages} {609}
  (\bibinfo {year} {1973})}\BibitemShut {NoStop}%
\bibitem [{\citenamefont {Acheson}\ and\ \citenamefont
  {Hide}(1973)}]{Ach73rev}%
  \BibitemOpen
  \bibfield  {author} {\bibinfo {author} {\bibfnamefont {D.~J.}\ \bibnamefont
  {Acheson}}\ and\ \bibinfo {author} {\bibfnamefont {R.}~\bibnamefont {Hide}},\
  }\href@noop {} {\bibfield  {journal} {\bibinfo  {journal} {Rep. Prog. Phys.}\
  }\textbf {\bibinfo {volume} {36}},\ \bibinfo {pages} {159} (\bibinfo {year}
  {1973})}\BibitemShut {NoStop}%
\bibitem [{\citenamefont {Balbus}\ and\ \citenamefont {Hawley}(1991)}]{Bal91}%
  \BibitemOpen
  \bibfield  {author} {\bibinfo {author} {\bibfnamefont {S.~A.}\ \bibnamefont
  {Balbus}}\ and\ \bibinfo {author} {\bibfnamefont {J.~F.}\ \bibnamefont
  {Hawley}},\ }\href@noop {} {\bibfield  {journal} {\bibinfo  {journal}
  {Astrophys. J.}\ }\textbf {\bibinfo {volume} {376}},\ \bibinfo {pages} {214}
  (\bibinfo {year} {1991})}\BibitemShut {NoStop}%
\bibitem [{\citenamefont {Thiffeault}\ and\ \citenamefont
  {Morrison}(1998)}]{Thi98}%
  \BibitemOpen
  \bibfield  {author} {\bibinfo {author} {\bibfnamefont {J.}~\bibnamefont
  {Thiffeault}}\ and\ \bibinfo {author} {\bibfnamefont {P.~J.}\ \bibnamefont
  {Morrison}},\ }\href@noop {} {\bibfield  {journal} {\bibinfo  {journal}
  {Annals of New York Academy of Sciences}\ }\textbf {\bibinfo {volume}
  {867}},\ \bibinfo {pages} {109} (\bibinfo {year} {1998})}\BibitemShut
  {NoStop}%
\bibitem [{\citenamefont {Andreussi}\ \emph {et~al.}(2010)\citenamefont
  {Andreussi}, \citenamefont {Morrison},\ and\ \citenamefont
  {Pegoraro}}]{amp0}%
  \BibitemOpen
  \bibfield  {author} {\bibinfo {author} {\bibfnamefont {T.}~\bibnamefont
  {Andreussi}}, \bibinfo {author} {\bibfnamefont {P.~J.}\ \bibnamefont
  {Morrison}}, \ and\ \bibinfo {author} {\bibfnamefont {F.}~\bibnamefont
  {Pegoraro}},\ }\href@noop {} {\bibfield  {journal} {\bibinfo  {journal}
  {Plasma Phys. Cont. Fusion}\ }\textbf {\bibinfo {volume} {52}},\ \bibinfo
  {pages} {055001} (\bibinfo {year} {2010})}\BibitemShut {NoStop}%
\bibitem [{\citenamefont {Andreussi}\ \emph {et~al.}(2012)\citenamefont
  {Andreussi}, \citenamefont {Morrison},\ and\ \citenamefont
  {Pegoraro}}]{amp1}%
  \BibitemOpen
  \bibfield  {author} {\bibinfo {author} {\bibfnamefont {T.}~\bibnamefont
  {Andreussi}}, \bibinfo {author} {\bibfnamefont {P.~J.}\ \bibnamefont
  {Morrison}}, \ and\ \bibinfo {author} {\bibfnamefont {F.}~\bibnamefont
  {Pegoraro}},\ }\href@noop {} {\bibfield  {journal} {\bibinfo  {journal}
  {Phys. Plasmas}\ }\textbf {\bibinfo {volume} {19}},\ \bibinfo {pages}
  {052102} (\bibinfo {year} {2012})}\BibitemShut {NoStop}%
\bibitem [{\citenamefont {Padhye}\ and\ \citenamefont
  {Morrison}(1996)}]{padhye}%
  \BibitemOpen
  \bibfield  {author} {\bibinfo {author} {\bibfnamefont {N.}~\bibnamefont
  {Padhye}}\ and\ \bibinfo {author} {\bibfnamefont {P.~J.}\ \bibnamefont
  {Morrison}},\ }\href@noop {} {\bibfield  {journal} {\bibinfo  {journal}
  {Phys. Lett. A}\ }\textbf {\bibinfo {volume} {219}},\ \bibinfo {pages} {287}
  (\bibinfo {year} {1996})}\BibitemShut {NoStop}%
\bibitem [{\citenamefont {Hazeltine}\ \emph {et~al.}(1985)\citenamefont
  {Hazeltine}, \citenamefont {Kotschenreuther},\ and\ \citenamefont
  {Morrison}}]{Haz85}%
  \BibitemOpen
  \bibfield  {author} {\bibinfo {author} {\bibfnamefont {R.~D.}\ \bibnamefont
  {Hazeltine}}, \bibinfo {author} {\bibfnamefont {M.}~\bibnamefont
  {Kotschenreuther}}, \ and\ \bibinfo {author} {\bibfnamefont {P.~J.}\
  \bibnamefont {Morrison}},\ }\href@noop {} {\bibfield  {journal} {\bibinfo
  {journal} {Phys. Fluids}\ }\textbf {\bibinfo {volume} {28}},\ \bibinfo
  {pages} {2466} (\bibinfo {year} {1985})}\BibitemShut {NoStop}%
\bibitem [{\citenamefont {Dagnelund}\ and\ \citenamefont
  {Pavlenko}(2005)}]{Dag05}%
  \BibitemOpen
  \bibfield  {author} {\bibinfo {author} {\bibfnamefont {D.}~\bibnamefont
  {Dagnelund}}\ and\ \bibinfo {author} {\bibfnamefont {V.~P.}\ \bibnamefont
  {Pavlenko}},\ }\href@noop {} {\bibfield  {journal} {\bibinfo  {journal}
  {Phys. Scripta}\ }\textbf {\bibinfo {volume} {71}},\ \bibinfo {pages} {293}
  (\bibinfo {year} {2005})}\BibitemShut {NoStop}%
\bibitem [{\citenamefont {Khater}\ \emph {et~al.}(2005)\citenamefont {Khater},
  \citenamefont {Moawad},\ and\ \citenamefont {Callebaut}}]{Kha05}%
  \BibitemOpen
  \bibfield  {author} {\bibinfo {author} {\bibfnamefont {A.~H.}\ \bibnamefont
  {Khater}}, \bibinfo {author} {\bibfnamefont {S.~M.}\ \bibnamefont {Moawad}},
  \ and\ \bibinfo {author} {\bibfnamefont {D.~K.}\ \bibnamefont {Callebaut}},\
  }\href@noop {} {\bibfield  {journal} {\bibinfo  {journal} {Phys. of Plasmas}\
  }\textbf {\bibinfo {volume} {12}},\ \bibinfo {pages} {012316} (\bibinfo
  {year} {2005})}\BibitemShut {NoStop}%
\bibitem [{\citenamefont {Tassi}\ and\ \citenamefont {Morrison}(2011)}]{Tas11}%
  \BibitemOpen
  \bibfield  {author} {\bibinfo {author} {\bibfnamefont {E.}~\bibnamefont
  {Tassi}}\ and\ \bibinfo {author} {\bibfnamefont {P.~J.}\ \bibnamefont
  {Morrison}},\ }\href@noop {} {\bibfield  {journal} {\bibinfo  {journal}
  {Phys. of Plasmas}\ }\textbf {\bibinfo {volume} {18}},\ \bibinfo {pages}
  {032115} (\bibinfo {year} {2011})}\BibitemShut {NoStop}%
\bibitem [{\citenamefont {Morrison}(1986)}]{morrison86}%
  \BibitemOpen
  \bibfield  {author} {\bibinfo {author} {\bibfnamefont {P.~J.}\ \bibnamefont
  {Morrison}},\ }\href@noop {} {\bibfield  {journal} {\bibinfo  {journal}
  {Bull. Am. Phys. Soc.}\ }\textbf {\bibinfo {volume} {31}},\ \bibinfo {pages}
  {1609} (\bibinfo {year} {1986})}\BibitemShut {NoStop}%
\bibitem [{\citenamefont {Throumoulopoulos}\ and\ \citenamefont
  {Tasso}(1999)}]{tasso}%
  \BibitemOpen
  \bibfield  {author} {\bibinfo {author} {\bibfnamefont {G.~N.}\ \bibnamefont
  {Throumoulopoulos}}\ and\ \bibinfo {author} {\bibfnamefont {H.}~\bibnamefont
  {Tasso}},\ }\href@noop {} {\bibfield  {journal} {\bibinfo  {journal} {J.
  Plasma Phys.}\ }\textbf {\bibinfo {volume} {62}},\ \bibinfo {pages} {449}
  (\bibinfo {year} {1999})}\BibitemShut {NoStop}%
\bibitem [{\citenamefont {Greene}\ and\ \citenamefont {Johnson}(1968)}]{Gre68}%
  \BibitemOpen
  \bibfield  {author} {\bibinfo {author} {\bibfnamefont {J.}~\bibnamefont
  {Greene}}\ and\ \bibinfo {author} {\bibfnamefont {J.~L.}\ \bibnamefont
  {Johnson}},\ }\href@noop {} {\bibfield  {journal} {\bibinfo  {journal}
  {Plasma Physics}\ }\textbf {\bibinfo {volume} {10}},\ \bibinfo {pages} {729}
  (\bibinfo {year} {1968})}\BibitemShut {NoStop}%
\bibitem [{\citenamefont {White}(2006)}]{Whi06}%
  \BibitemOpen
  \bibfield  {author} {\bibinfo {author} {\bibfnamefont {R.~B.}\ \bibnamefont
  {White}},\ }\href@noop {} {\emph {\bibinfo {title} {{The Theory of Toroidally
  Confined Plasmas}}}}\ (\bibinfo  {publisher} {Imperial College Press,
  London},\ \bibinfo {year} {2006})\BibitemShut {NoStop}%
\bibitem [{\citenamefont {Julien}\ and\ \citenamefont
  {Knobloch}(2010)}]{Jul10}%
  \BibitemOpen
  \bibfield  {author} {\bibinfo {author} {\bibfnamefont {K.}~\bibnamefont
  {Julien}}\ and\ \bibinfo {author} {\bibfnamefont {E.}~\bibnamefont
  {Knobloch}},\ }\href@noop {} {\bibfield  {journal} {\bibinfo  {journal}
  {Phil. Trans. R. Soc. A}\ }\textbf {\bibinfo {volume} {368}},\ \bibinfo
  {pages} {1607} (\bibinfo {year} {2010})}\BibitemShut {NoStop}%
\end{thebibliography}

%%%%%%%%%%%%%%%%%%%%%%%%%%%%%%%%%%%%
%%%%%%%%%%%%%%%%%%%%%%%%%%%%%%%%%%%%

%

\end{document}